\documentclass[preprint,review,12pt]{elsarticle}

\usepackage{graphicx}
\usepackage{amssymb} 
\usepackage{amsthm} 
\usepackage{lineno}

\journal{Nuclear Physics A} 

\begin{document}

\begin{frontmatter} 

\title{Highlights from STAR}

\author{Xin Dong (for the STAR\fnref{col1} Collaboration)}
\fntext[col1] {A list of members of the STAR Collaboration and acknowledgements can be found at the end of this issue.}
\address{Lawrence Berkeley National Laboratory \\ MS70R0319, 1 Cyclotron Road, Berkeley, CA 94720, USA}


\begin{abstract} 
In these proceedings, I highlight some selected results from the STAR experiment that were presented in the Quark Matter 2012 conference.
\end{abstract} 

\end{frontmatter} 

\linenumbers

\section{Introduction}

The physics program of the STAR experiment covers various aspects of QCD frontiers. Diverse heavy ion collisions provided by the RHIC machine allow us to systematically study the properties of the strongly-coupled Quark Gluon Plasma (sQGP) at the top RHIC energy, as well as to map out the phase structure of QCD matter by scanning the phase diagram with variable collision energies. The $p$($d$)-A collisions provide opportunities to control the cold nuclear effects and to study the QCD in the low-$x$ regime.

\begin{figure}[htbp]
\centerline{
\includegraphics[width=0.45\textwidth] {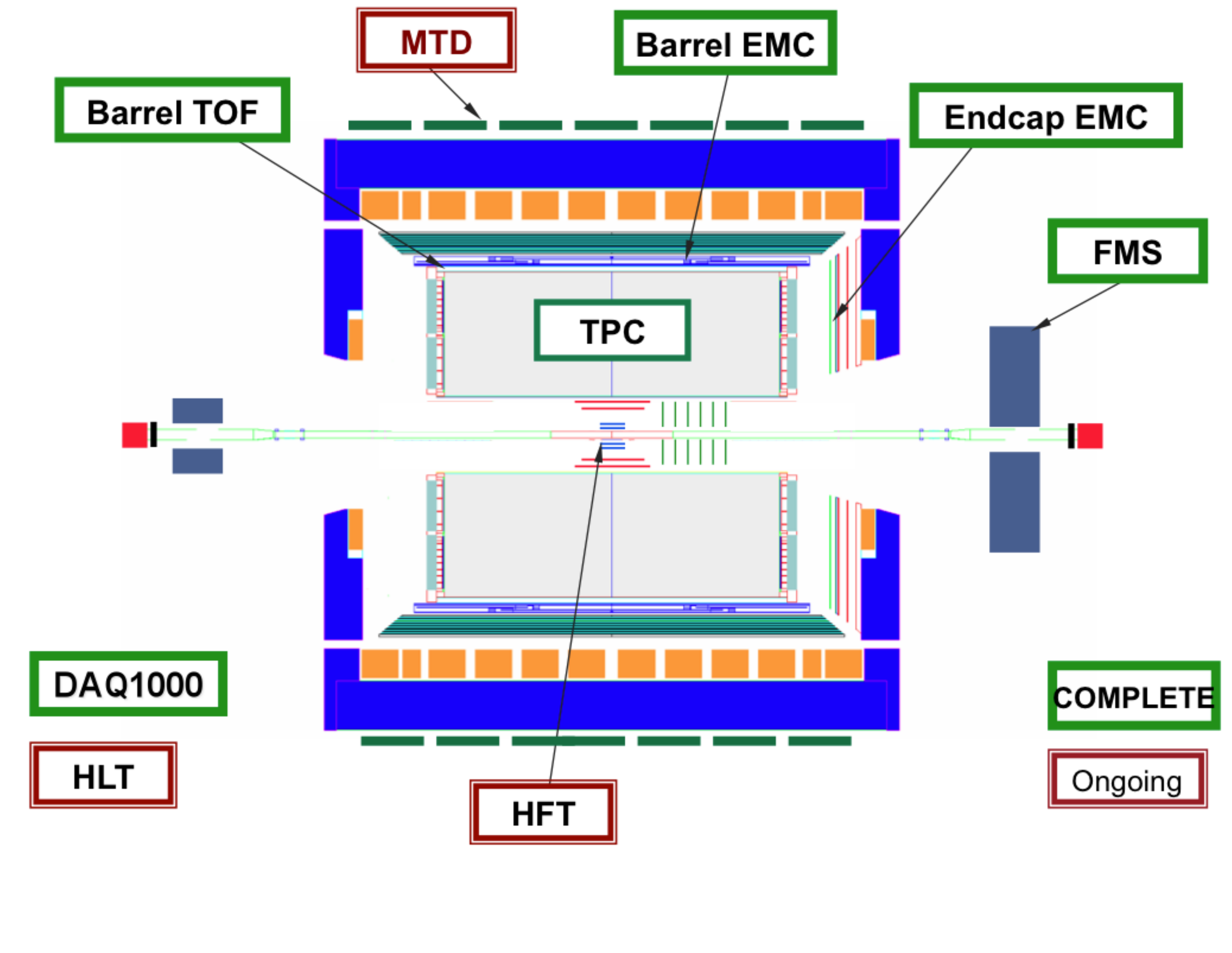}
\hspace{0.3in}
\includegraphics[width=0.5\textwidth] {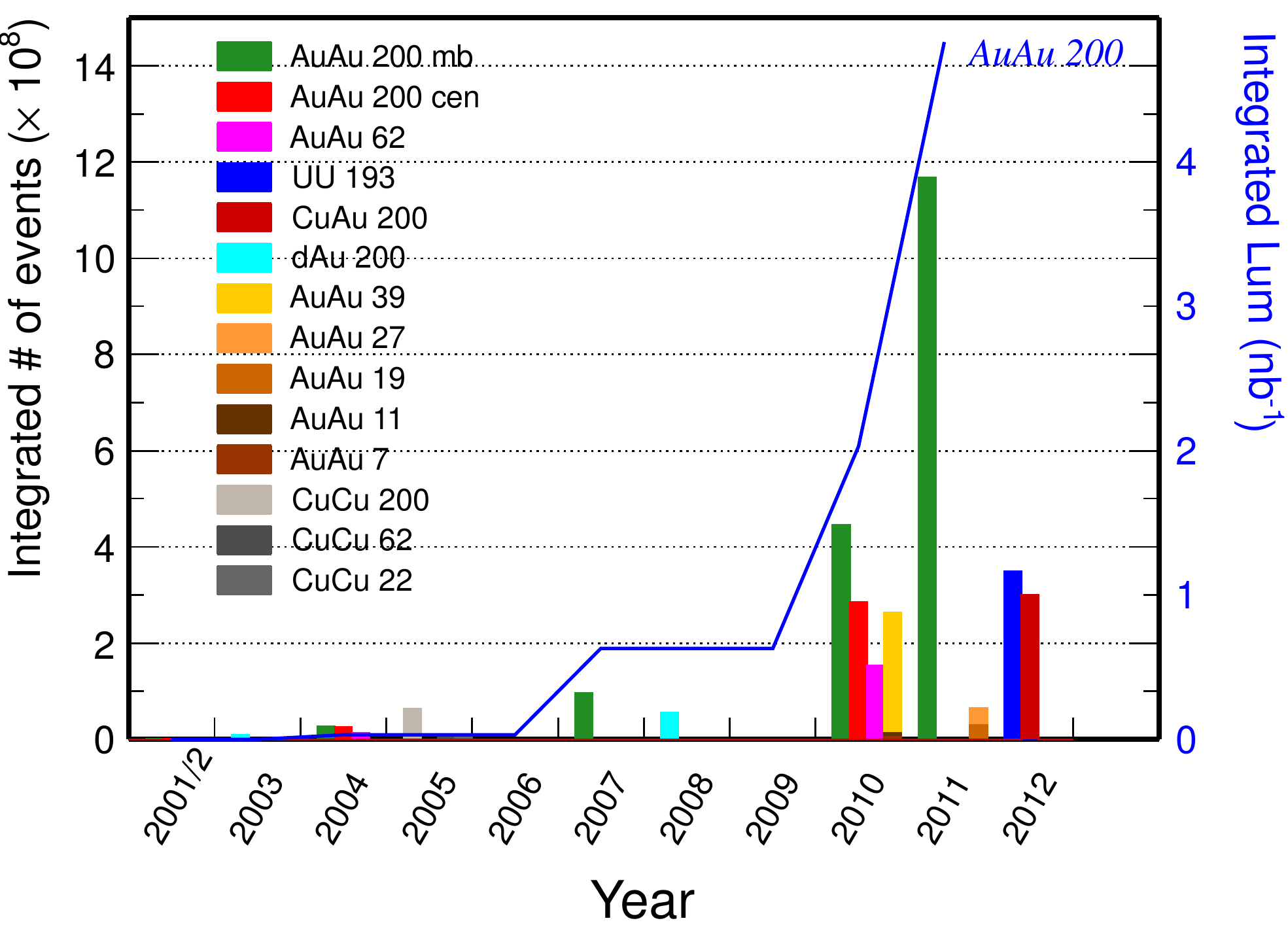}}
\caption[]{(Left): A side view of the STAR detector and its main subsystems. (Right): Heavy-ion minimum bias/central data sets (histograms) and integrated luminosity (line) recorded by the STAR detector. }
\label{detdata}
\end{figure}

The STAR detector at RHIC is well suited to carry out these physics programs~\cite{STARdet}. A side view of the STAR detector and its main subsystems are shown in the left plot of Fig.~\ref{detdata}. The STAR detector has a large and uniform acceptance with excellent particle identification at mid-rapidity ($|\eta|<\sim1$) across all collider energies. The calorimeter subsystems extend to forward rapidity (up to $\eta\sim4$) enabling to probe extreme kinematic regions. Figure~\ref{detdata} right panel shows the recorded heavy ion data samples from all RHIC runs by the STAR detector. Significant amounts of data have been accumulated since 2010 when the fast data acquisition and the TOF subsystem upgrades were completed. STAR has entered the era of precision measurements to study the QCD in hot and cold nuclear matter.

In these proceedings, I will highlight some selected STAR results that were presented in the Quark Matter 2012 conference. Please refer to the STAR contributed articles in these proceedings for more details.


\section{Forward dihadron correlations to search for the Color Glass Condensate (CGC)}

The gluon density within a hadron increases drastically with decreasing $x$, but will saturate when gluon recombination balances gluon splitting and the collinear factorization fails. This idea has been casted in a precise theoretical frame known as Color Glass Condensate (CGC) effective theory model~\cite{CGC}. The saturation scale ($Q_s^2$) at a given $x$ increases as $A^{1/3}$, thus it becomes more effective to search for this phenomena using collisions with nuclei. At the leading order 2$\rightarrow$2 process, triggering the produced particle in the forward region and varying the associated particle $\eta$ regions offers an opportunity to study the $x$ dependence of particle production enabling to probe regimes from dilute parton gas to possible CGC. Previously, STAR has reported the measurement of forward $\pi^0$-$\pi^0$ correlations in the rapidity region $2.5<\eta<4$ with the FMS detector. The back-to-back correlation in $d$ + Au collisions is significantly suppressed or broadened compared to that in $p+p$ collisions, consistent with the CGC expectation~\cite{dAuforwardCorr}. The measurement of forward ($2.5<\eta<4$) and mid rapidity ($-1<\eta<1$) $\pi^0$ correlations show no significant difference in the correlation widths between $p+p$ and $d$ + Au collisions~\cite{dAuforwardCorr}.

\begin{figure}[ht]
\centerline{
\includegraphics[width=0.45\textwidth] {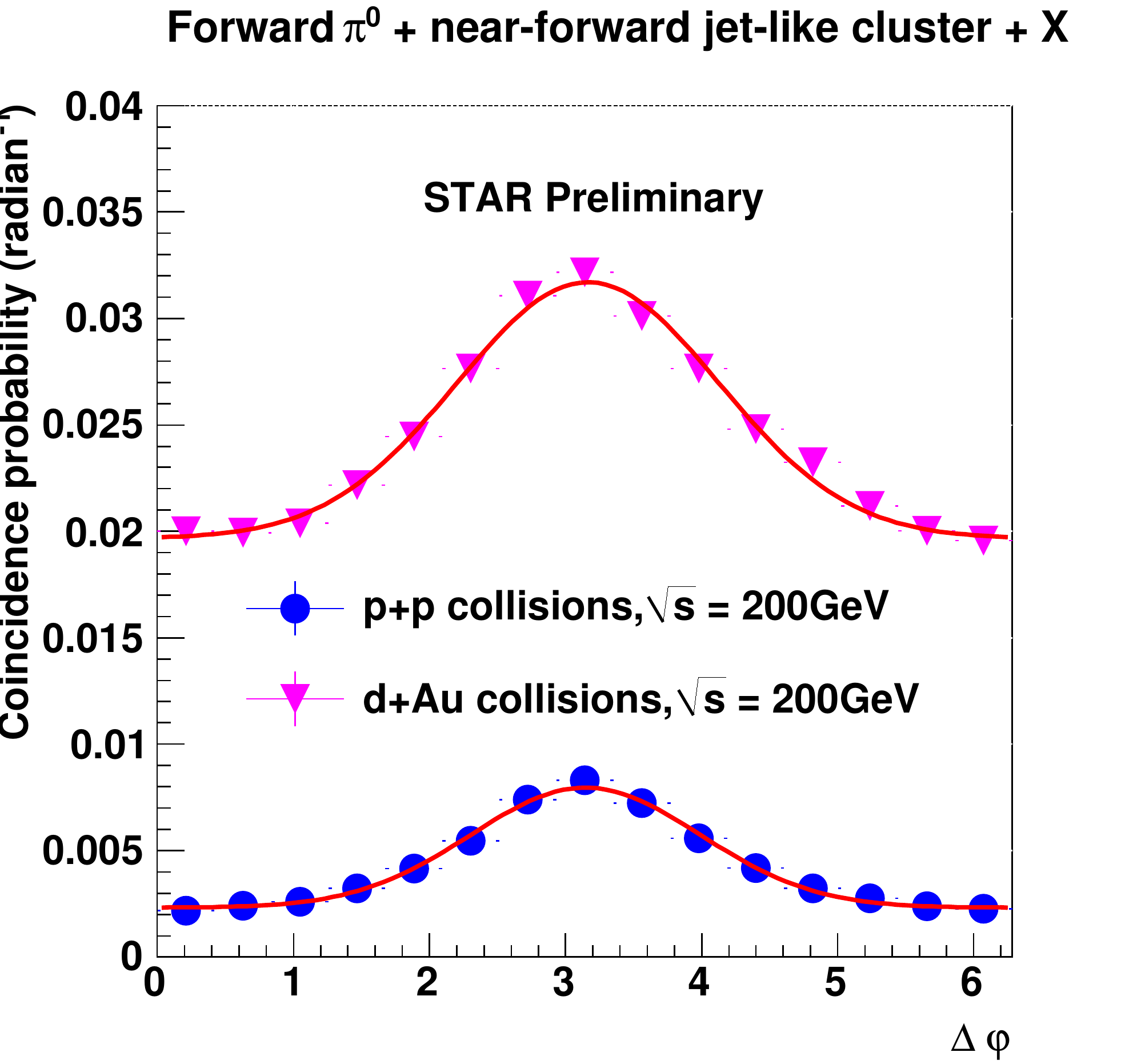}
\hspace{0.3in}
\includegraphics[width=0.45\textwidth] {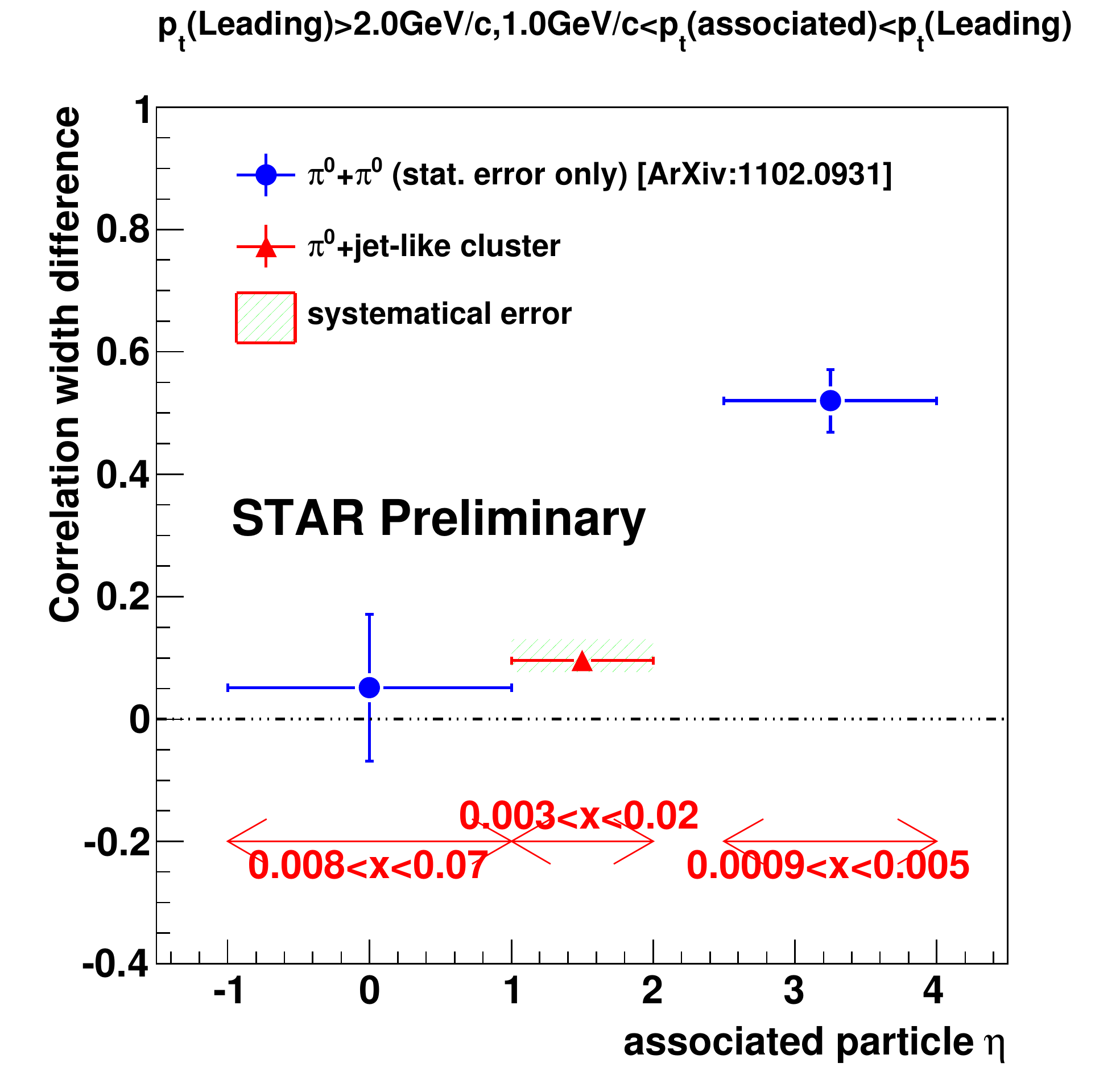}}
\caption[]{(Left): Dihadron azimuthal angle difference $\Delta\phi$ distributions between forward $\pi^0$ triggered by FMS (2.5$<\eta<$4 and near-forward jet-like cluster from EEMC (1$<\eta<$2) in $p+p$ and $d$ + Au collisions at $\sqrt{s_{NN}}$ = 200 GeV. (Right): Correlation width difference between $p+p$ and $d$ + Au triggered by the FMS $\pi^0$ vs. the associated particle $\eta$. }
\label{dihardon}
\end{figure}

In this conference, we report the dihadron correlation measurement with forward $\pi^0$ from FMS and near-forward (1$<\eta<$2) jet-like cluster from EEMC~\cite{Li}. It is sensitive to the intermediate $x$ region to study the transition between previous forward-forward and forward-midrapidity correlations. The correlation distributions in $p+p$ and $d$ + Au collisions at $\sqrt{s_{NN}}$ = 200 GeV are shown in the left plot of Fig.~\ref{dihardon}. Red lines depict gaussian fits to two distributions to extract the widths. The width difference between $d$ + Au and $p+p$ from forward triggered correlations with associated particles coming from different $\eta$ regions is summarized in the right plot of Fig.~\ref{dihardon}. Also shown in the plot are the corresponding gluon $x$ regions in the Au nuclei for correlations with different associate particle $\eta$. The plot shows that with increasing associate particle $\eta$ - and thus smaller gluon $x$ - the correlation width difference increases. The $x$ dependence feature in forward dihardon correlations is consistent with a smooth transition from the dilute parton gas to the CGC regime.
 
\section{Towards precision understanding of the sQGP properties}

\begin{figure}[htbp]
\centerline{
\includegraphics[width=1.0\textwidth]{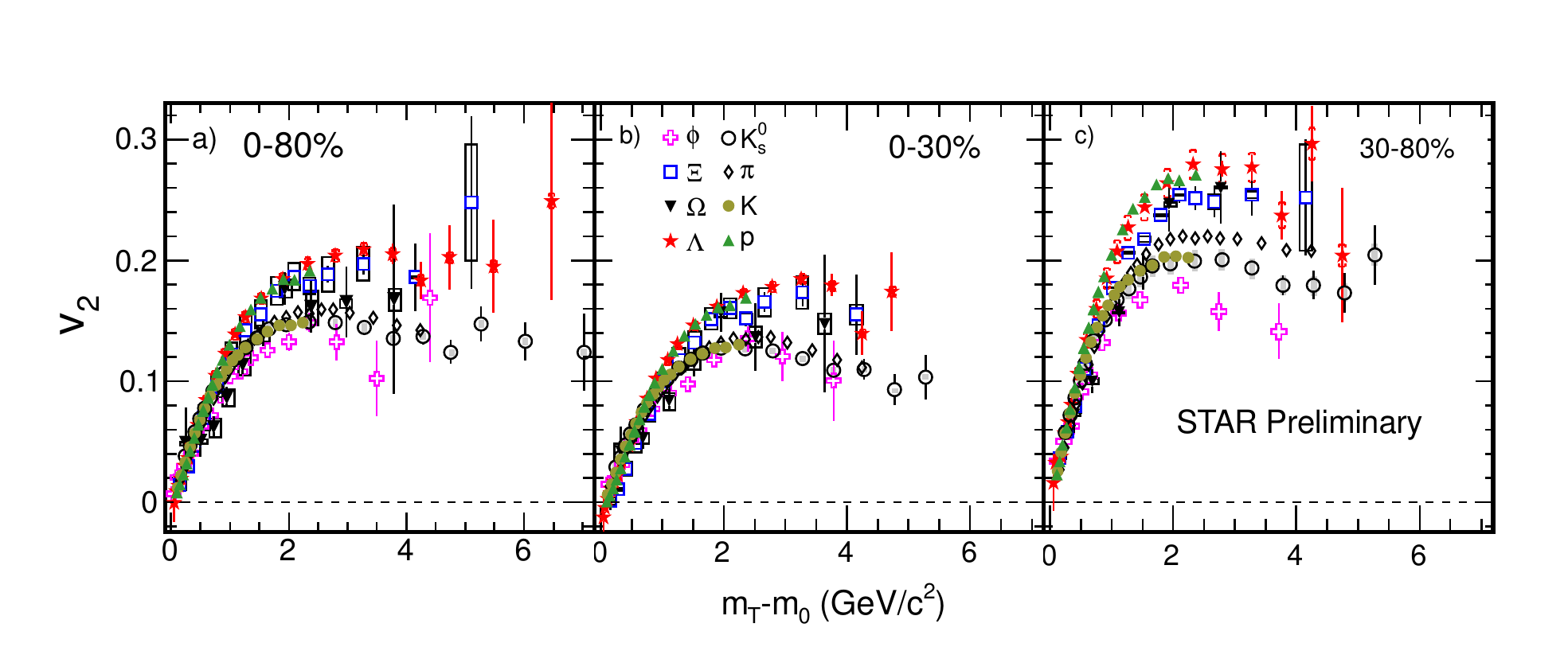}}
\caption[]{Identified particle $v_2$ vs. transverse kinetic energy ($m_T-m_0$) in Au + Au 200 GeV for 0-80\% minimum bias (a), 0-30\% (b) and 30-80\% (c) centrality bins. Error bars on the data points are statistical, and caps, open or shaded boxes are systematic uncertainties on different particles for better illustration.}
\label{PIDv2_200GeV}
\end{figure}

The strong elliptic flow ($v_2$) and the number of constituent quark (NCQ) scaling found for multi-strange hadrons at RHIC top energy are clear evidence of partonic collectivity, a critical feature of the sQGP~\cite{partonicFlow}.  With the unprecedented statistics collected in the past years, STAR is able to make systematic measurements to study the partonic collectivity and to investigate the sQGP properties~\cite{Nasim}. Figure~\ref{PIDv2_200GeV} shows the recent $v_2$ measurements of identified particles ($\pi^{\pm}$,$K^{\pm}$,$K_{S}^0$,$p$,$\bar{p}$,$\phi$,$\Lambda$,$\bar\Lambda$,$\Xi^{-}$,$\bar\Xi^{+}$,$\Omega^{-}$,$\bar\Omega^{+}$) as a function of transverse kinetic energy $m_T-m_0$  from Au + Au 200 GeV minimum bias (left), 0-30\% (middle) and 30-80\% (right) centrality collisions, respectively. In minimum bias and 0-30\% centrality, the measurements show a clear baryon/meson grouping and the NCQ scaling holds within 10\% for all particles at $(m_T-m_0)/n_q>0.6$ GeV/$c^2$, suggesting that partonic collectivity dominates the final observed $v_2$. While in 30-80\% centrality, one can see the baryon/meson grouping starts to collapse, and the $v_2/n_q$ of multi-strange hadrons ($\phi$,$\Xi$) deviates from that of $K_S^0$ beyond 10-15\%, suggesting smaller contributions from the partonic phase to the final collectivity. We also report systematic measurements of all hadronic flows for charged hadrons up to $n=5$~\cite{Pandit}. All these systematic measurements provide significant inputs to constrain the sQGP properties when comparing to hydrodynamic model calculations.

\begin{figure}[htbp]
\centerline{
\includegraphics[width=0.48\textwidth] {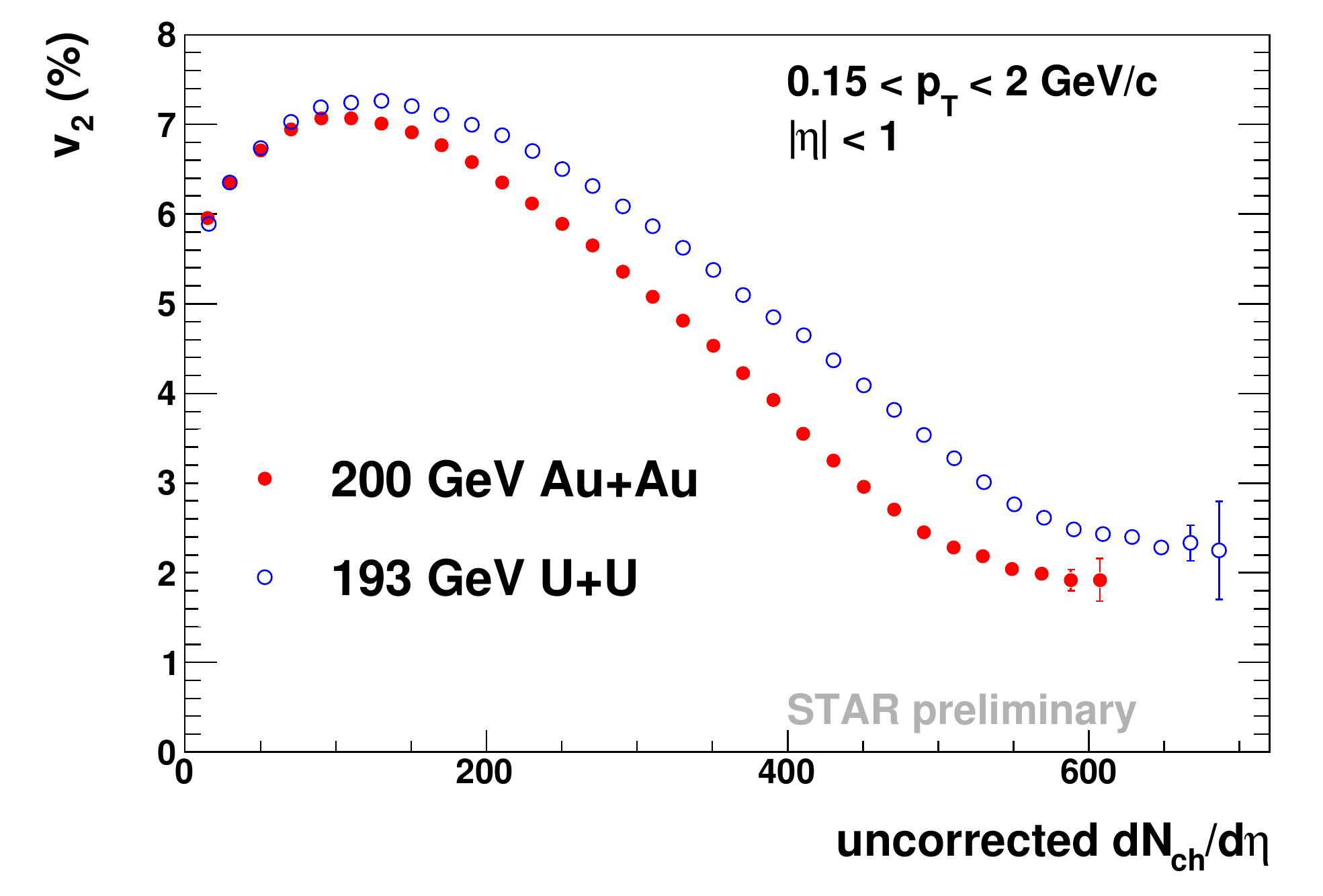}
\hspace{0.2in}
\includegraphics[width=0.48\textwidth]{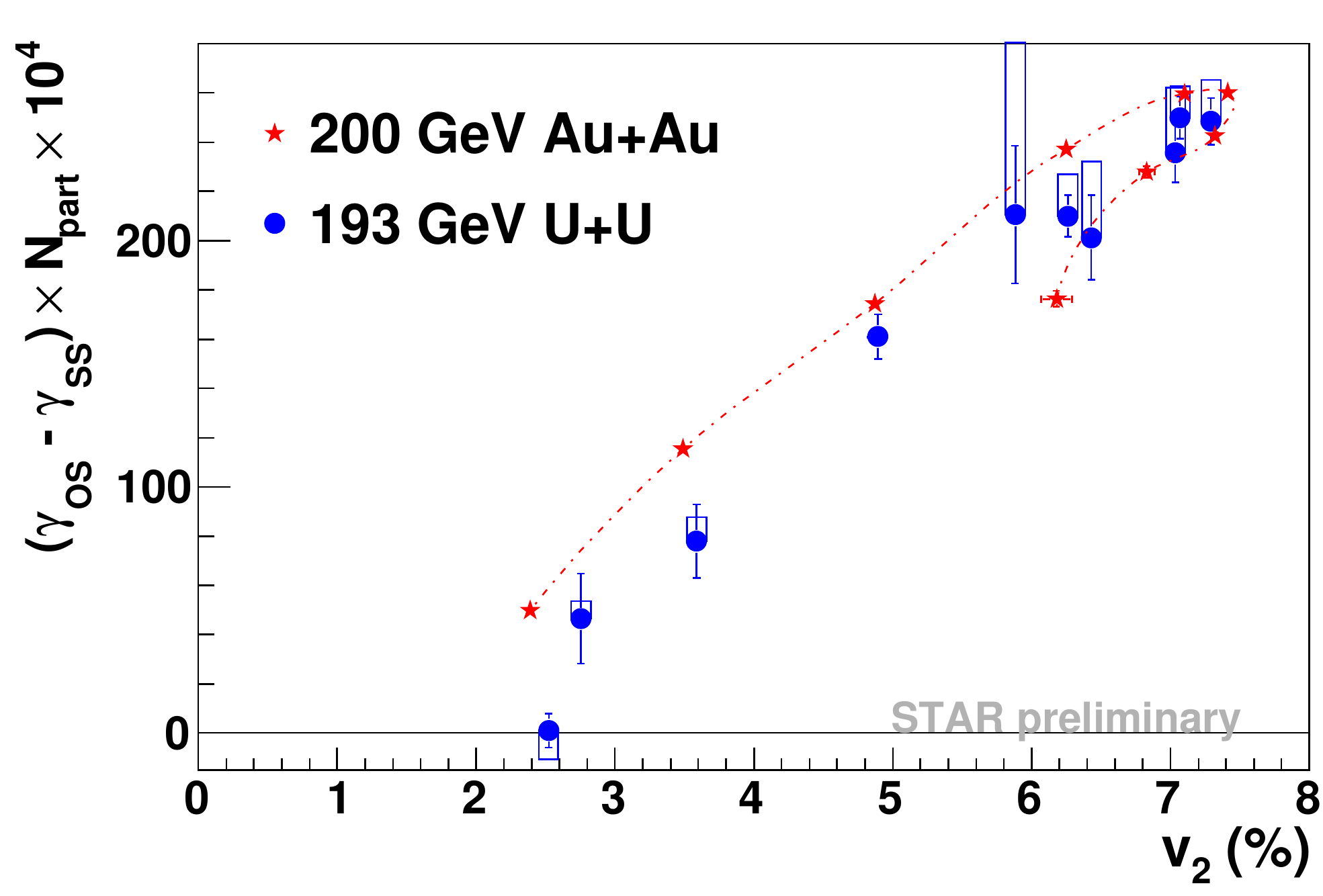}}
\caption[]{(Left) Charged hadron $v_2$  vs. uncorrected $dN_{ch}/d\eta$ for Au + Au 200 GeV and U + U 193 GeV collisions. (Right) Charge separation signal w.r.t. the event plane ($\gamma_{OS}-\gamma_{SS}$) from Au + Au (red) and U + U collisions (blue). The most left blue data point was obtained from the 0-1\% central triggered U + U collisions. Boxes depict the systematic uncertainties due to the possible pileup background source.}
\label{UUv2LPV}
\end{figure}

In 2012, RHIC for the first time collided two uranium beams. The uranium nucleus has a much larger mass and is largely deformed. It is very interesting to study the sQGP properties at higher particle density as well as with different collision orientations. Figure~\ref{UUv2LPV} left plot shows the integrated charged hadron $v_2$ using the $\eta$-sub event plane method vs. uncorrected charged particle density $dN_{ch}/d\eta$ in the region of $0.15<p_T<2$ GeV/$c$ and $|\eta|<1$ from 200 GeV Au + Au and 193 GeV U + U collisions~\cite{Wang}. One can see in the central and mid-central collisions, the $v_2$ in U + U collisions is always higher than that in Au + Au collisions at the same uncorrected $dN_{ch}/d\eta$.

STAR previously reported the measurement of the charge separation with respect to the event plane, which was motivated by the search for the parity violation in strong interactions and the chiral magnetic effect (CME) in heavy ion collisions~\cite{LPV}. There was some discussion on the physics background caused by the local energy/momentum conservation convoluted with the event-by-event elliptic flow~\cite{Pratt}. U + U collisions provide an opportunity to test the validation of the CME signal with tuning the magnitudes in $v_2$ and the magnetic field by different collision orientations~\cite{Voloshin}. Figure~\ref{UUv2LPV} right plot shows the charge separation signal w.r.t. the event plane (difference between $\gamma\equiv\langle\cos(\phi_a+\phi_b-2\Psi_{\rm{RP}})\rangle$ for opposite sign and same sign pairs) vs. the measured $v_2$ in Au + Au and U + U collisions. One particular note is that the most left data point in U + U collisions was obtained from the 0-1\% central triggered data sample where the magnetic field is expected to be significantly suppressed. The result is consistent with zero while there is still a sizable $v_2$ ($\sim$ 2.5\%) in 0-1\% central collisions. The observation is consistent with no signal when the magnetic field is turned off, which is expected from the CME~\cite{Wang}.

\begin{figure}[htbp]
\centerline{
\includegraphics[width=0.45\textwidth]{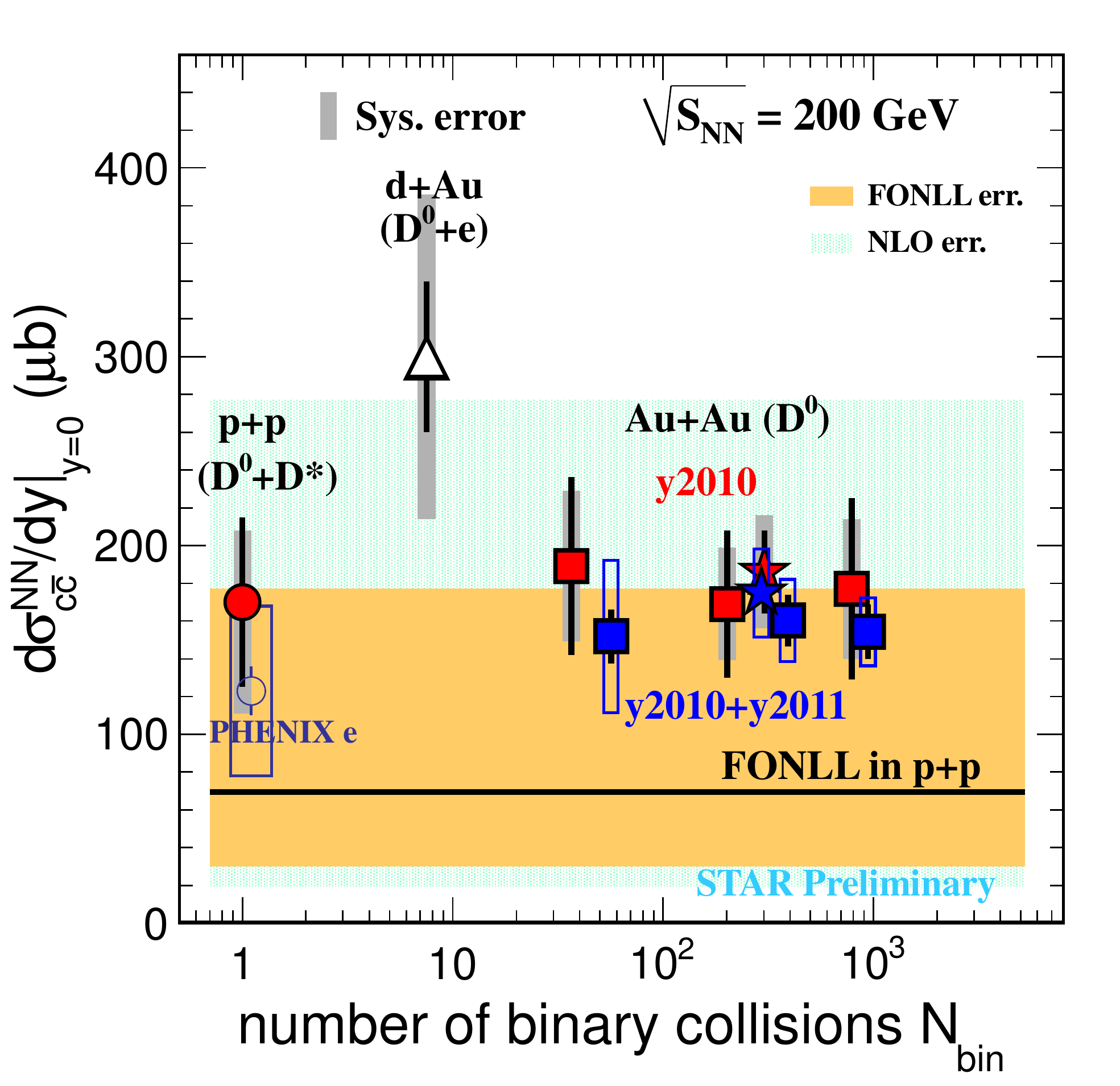}
\hspace{0.3in}
\includegraphics[width=0.52\textwidth]{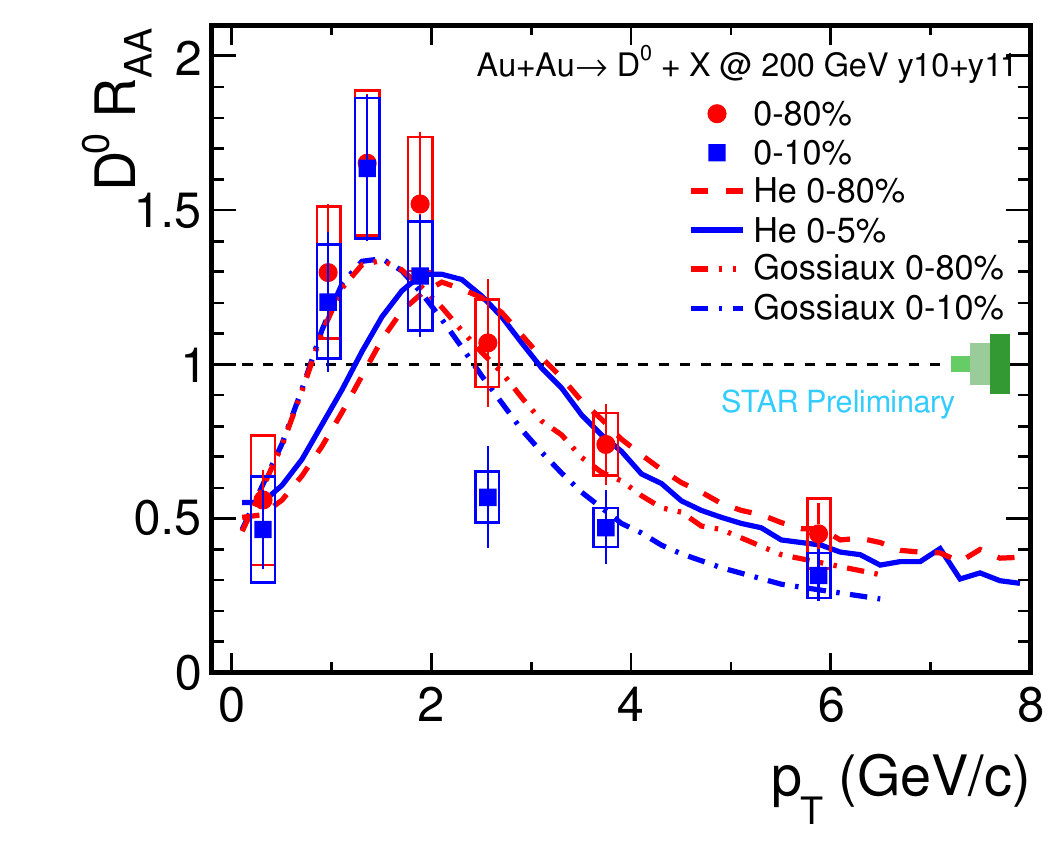}}
\caption[]{(Left): Charm production cross section per nucleon-nucleon collision from $p+p$ to central Au + Au collisions at 200 GeV; (Right): The $D^0$ meson $R_{AA}$ vs. $p_T$ for 0-10\% and 0-80\% centralities in Au + Au collisions at $\sqrt{s_{NN}}$ = 200 GeV compared to two transport model calculations. Error bars on the data points are statistical while boxes are systematic uncertainties.}
\label{D0AuAu}
\end{figure}

Heavy flavors are ideal probes to study the sQGP properties because of higher sensitivity due to their large masses. STAR previously reported open charm hadron measurements via hadronic decay channels in $p+p$ and Au + Au collisions at 200 GeV~\cite{Zhang}. With significantly improved statistics from year 2011 Au + Au collisions, we are able to carry out more systematic measurements on the charm total cross section as well as the $D^0$ $R_{AA}$~\cite{Tlusty}. Figure~\ref{D0AuAu} left plot shows the charm production cross section per nucleon-nucleon collision from $p+p$ to central Au + Au collisions. The data points from Au + Au collisions were extracted from the measurement of $D^0$ mesons assuming the same fragmentation ratio ($c\rightarrow D^0$) as in $p+p$. The results, with significantly improved precision, still show the charm total cross section follows the binary scaling from $p+p$ to central Au + Au collisions.

Figure~\ref{D0AuAu} right plot shows the $R_{AA}$ of $D^0$ mesons in central and minimum bias Au + Au collisions covering $0<p_T<\sim6$ GeV/$c$. Two important features: the data points show a modified hump structure at about $1-3$ GeV/$c$, which may indicate strong interactions between charm quarks with the medium; at $p_T>$ 3 GeV/$c$, there is a strong suppression in the observed $D^0$ $R_{AA}$, and the suppression is larger in central collisions, indicating large energy loss for the energetic charm quarks traversing through the medium. Also shown on the plot are two transport model calculations which reasonably describe the measured data points~\cite{CharmModels}.

\begin{figure}[htbp]
\centerline{
\includegraphics[width=0.88\textwidth]{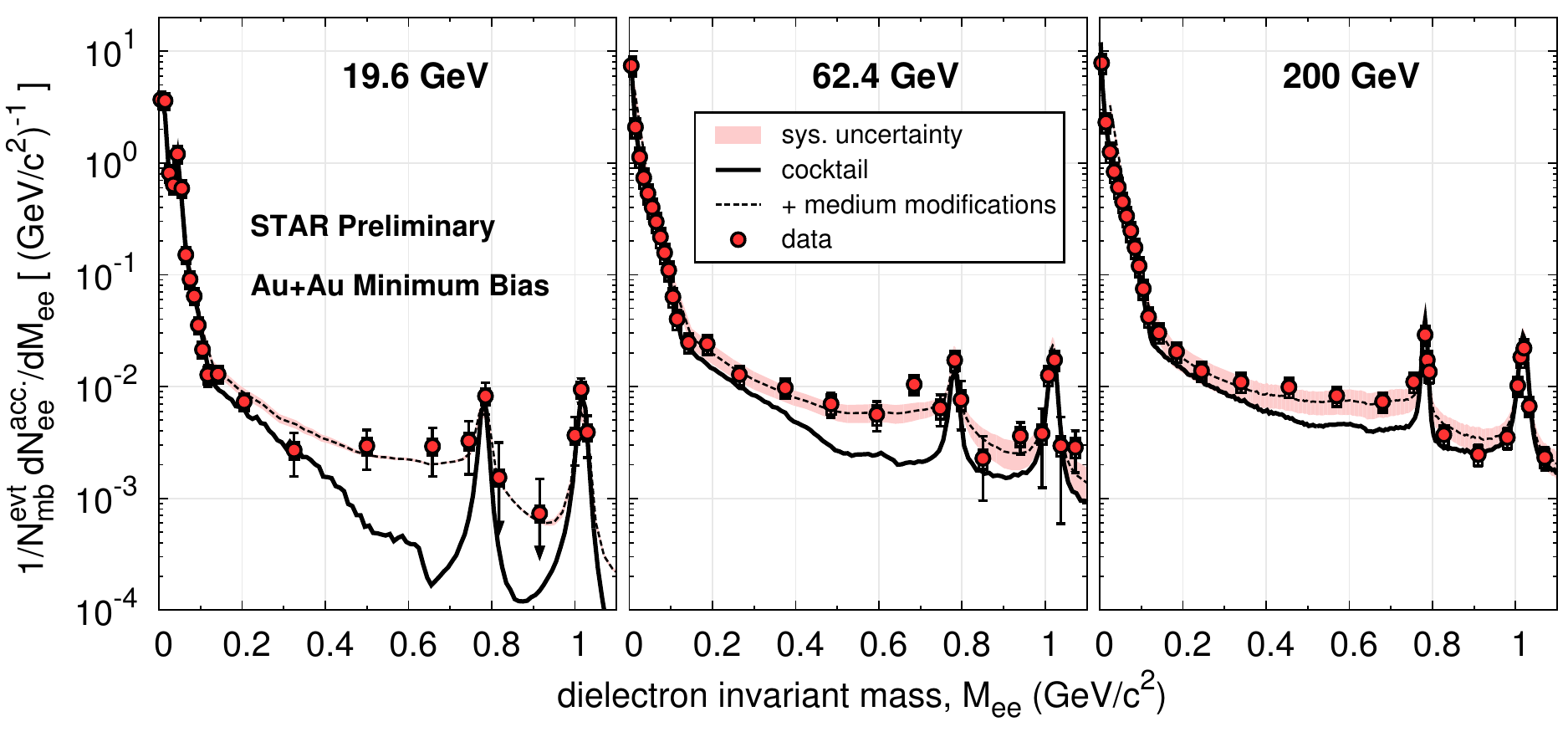}}
\caption[]{Dielectron mass spectra from Au + Au collisions at $\sqrt{s_{NN}}$ = 19.6, 62.4 and 200 GeV compared to the hadronic cocktail plus the medium vector meson and QGP contributions from model calculations. Uncertainties on the data points include statistical (bars) and systematical (boxes) and the red bands depict the uncertainties of cocktails.}
\label{dielectron}
\end{figure}

Dilepton pairs produced from the fireball carry clean information and offer insights into medium properties at different stages~\cite{Rapp}. STAR presented the dielectron production in 200 GeV Au + Au collisions at QM11, and reported a sizable enhancement compared to the hadronic cocktail in the mass region of 0.3-0.7 GeV/$c^2$~\cite{Zhao}. The excess can be described by a broadened $\rho$ in the hadronic medium~\cite{Rapp,PHSDandXu}. We have continued with more differential measurements and have studied the $p_T$ and centrality dependence as well as elliptic flow at 200 GeV~\cite{BHuang}. We also carried out a systematic measurement of the dielectron yield at various beam energies (19.6, 39, 62.4 GeV). Figure~\ref{dielectron} shows the dielectron mass spectrum within the STAR acceptance in Au + Au collisions at $\sqrt{s_{NN}}$ = 19.6, 62.4 and 200 GeV. The data show that the enhancement in the low mass region compared to the hadronic cocktail persists in all energies. Also included in the plot are calculated contributions of in-medium $\rho$ together with the QGP radiation from~\cite{Rapp}. The low mass excess can be consistently accounted for by the broadened $\rho$ production in the hadronic medium from 19.6 to 200 GeV.

\section{Results from Beam Energy Scan (BES) program}

To study the QCD phase structure and locate the first-order phase transition as well as a possible critical point~\cite{BESproposal}, RHIC has started the Beam Energy Scan (BES) program and the phase-I data have been taken in 2010 and 2011 with Au + Au collisions at energies of 39, 27, 19.6, 11.5, and 7.7 GeV. 

\begin{figure}[htbp]
\centerline{
\includegraphics[width=0.42\textwidth] {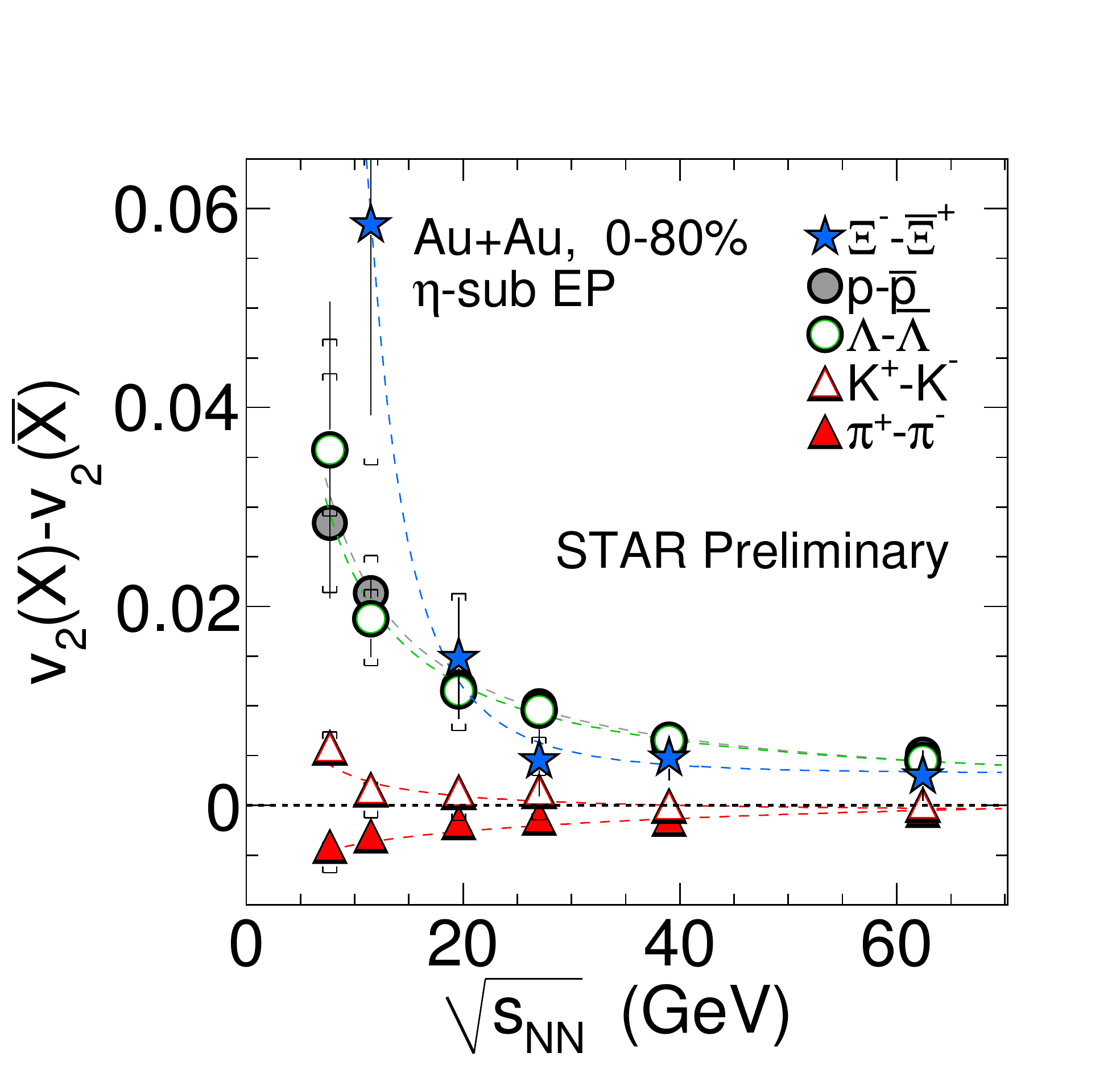}
\hspace{0.3in}
\includegraphics[width=0.52\textwidth]{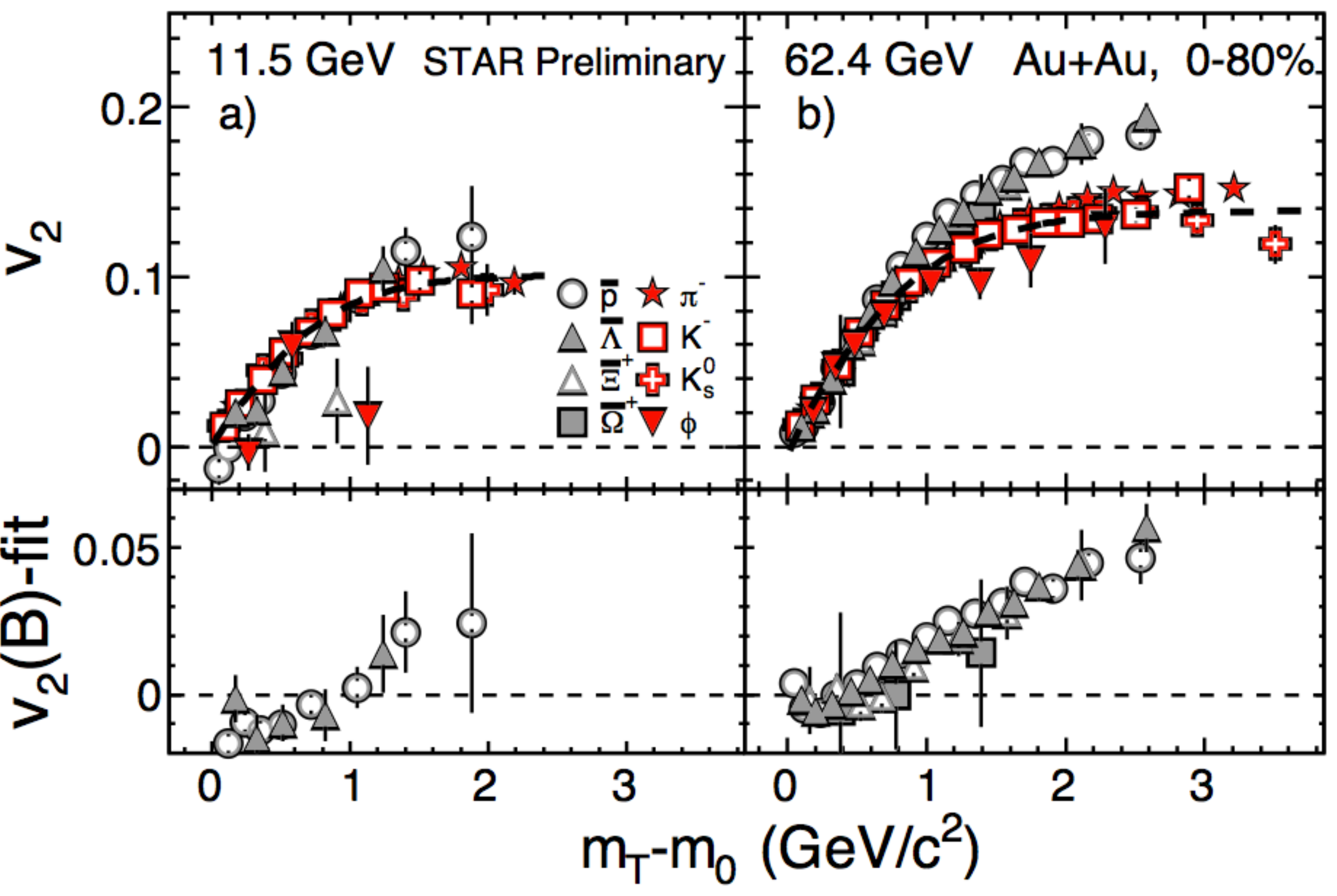}}
\caption[]{(Left): Particle and anti-particle $v_2$ difference vs. collision energy. Uncertainties on data points are statistical (bars) and systematical (brackets). (Right): Identified particle $v_2$ vs. $m_T-m_0$ for anti-particles in Au + Au collisions at 11.5 GeV and 62.4 GeV. Error bars are statistical only.}
\label{BESv2}
\end{figure}

One of the first questions when we lower the collision energies is whether these key sQGP signatures observed at the top RHIC energy turn off. Elliptic flow of identified particles and the NCQ scaling test have been proposed to address this question. We have made systematic measurements for identified particle $v_2$ in all beam energies~\cite{Shi}. Strikingly, we observed that the difference in $v_2$ between particles and anti-particles, which is negligible at top RHIC energy, starts to increase with decreasing collision energy, shown in the left plot of Fig.~\ref{BESv2}. The difference for baryons is larger than mesons. This suggests the breakdown of the NCQ scaling between particles and anti-particles at lower energies. In addition, we also observe that for anti-particles, the baryon/meson grouping at intermediate $p_T$ starts to collapse at 11.5 GeV, shown on the right plot of Fig.~\ref{BESv2}. These observations indicate that hadronic interactions become more dominant at lower beam energies.
We also observed changes of other key sQGP features at lower energies, e.g., the disappearance of high $p_T$ suppression~\cite{Sangaline,XZhang} and the charge separation signal~\cite{Wang}.

\begin{figure}[htbp]
\begin{minipage}[b]{0.45\textwidth} \centering\mbox{
\includegraphics[width=0.9\textwidth]{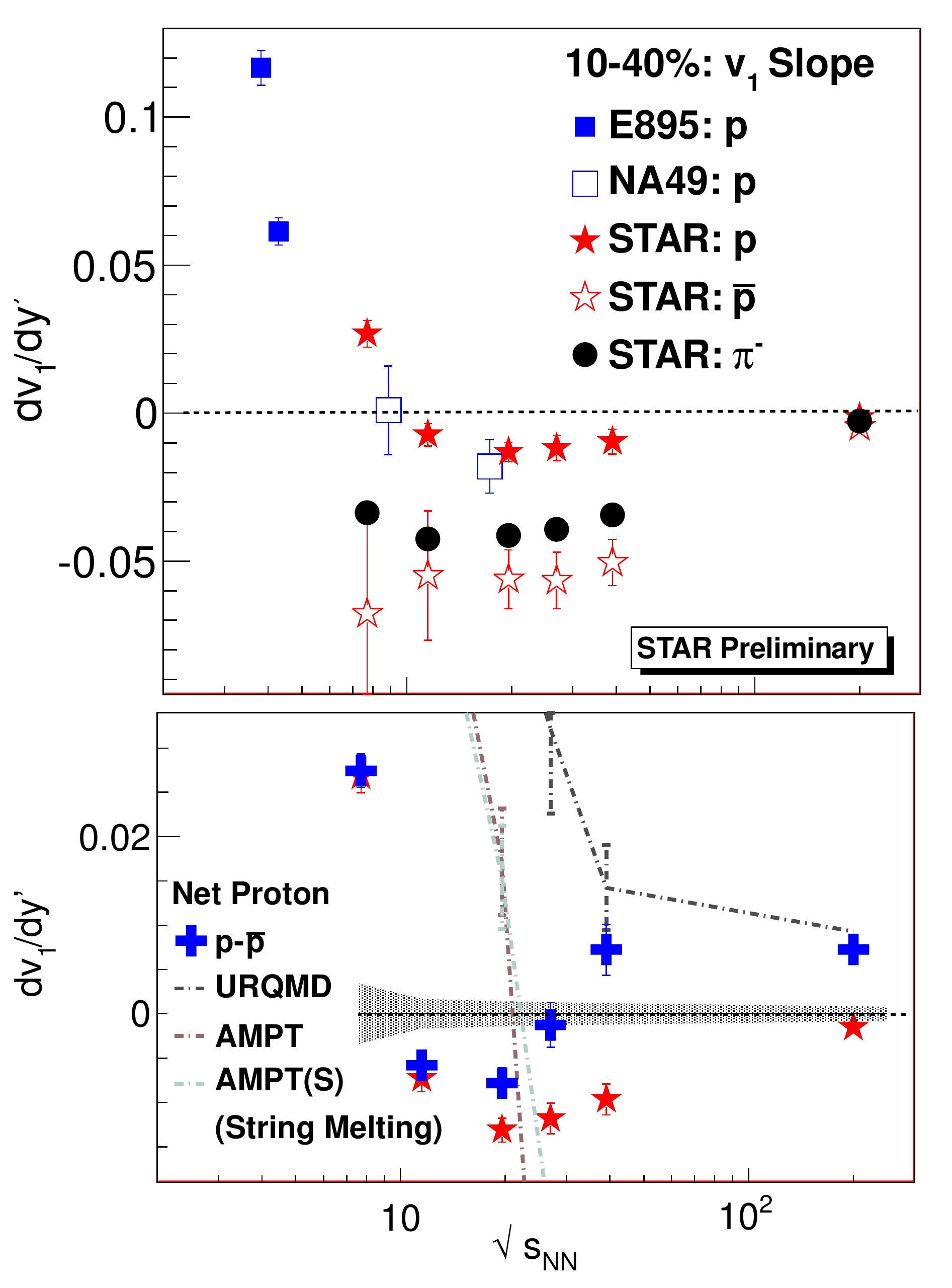}}
\caption[]{(Upper) Directed flow $v_1$ slope parameter of $p$, $\bar{p}$ and $\pi^{-}$ vs. energy. (Lower) net-proton $v_1$ slope parameter vs. energy and compared to transport model calculations. Error bars are statistical only.}
\label{proton_v1slope}
\end{minipage}
\hspace{0.3in}
\begin{minipage}[b]{0.52\textwidth} \centering\mbox{
\includegraphics[width=1.0\textwidth]{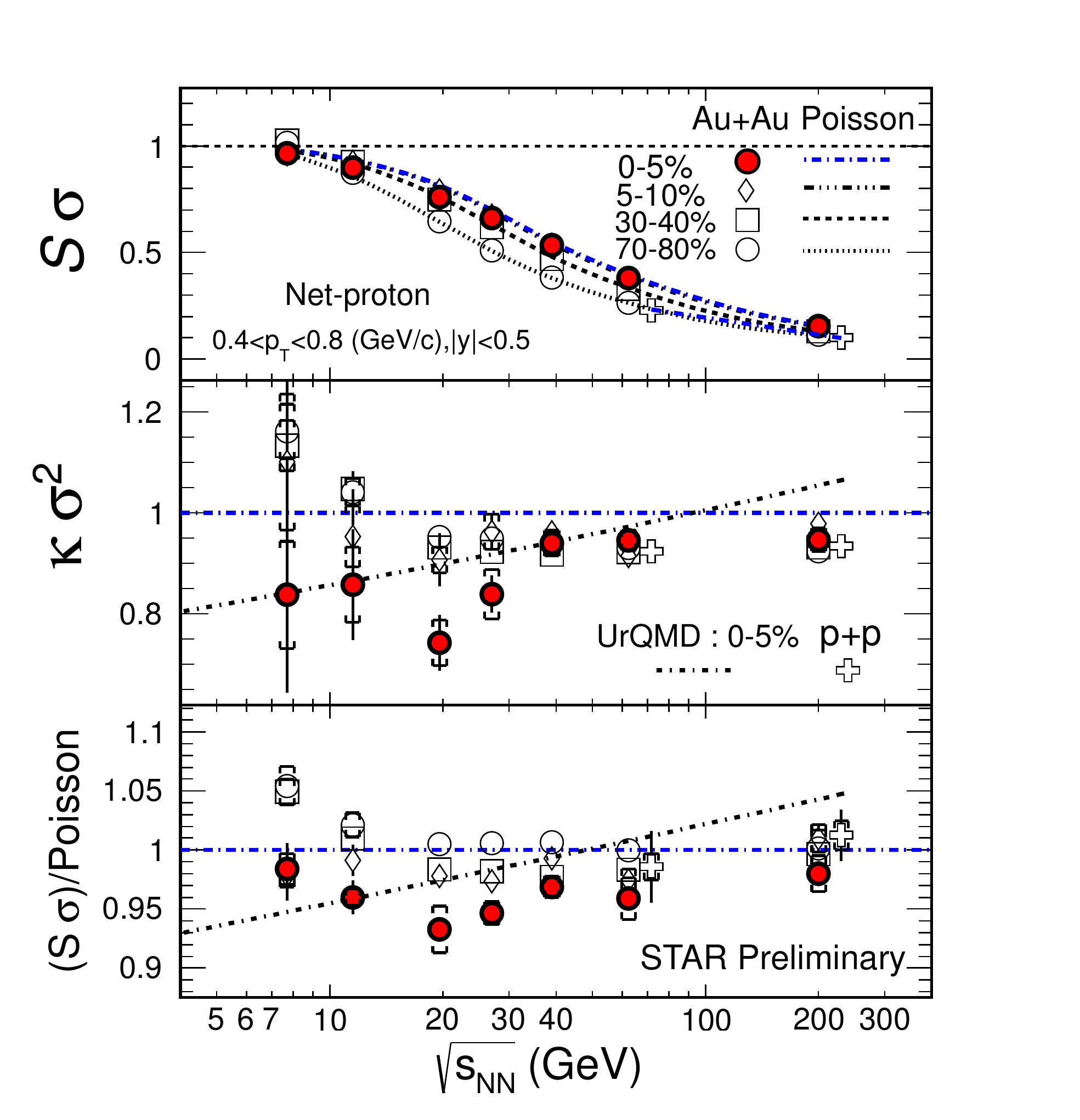}}
\caption[]{Moment products $S\sigma$, $\kappa\sigma^2$ of net-proton multiplicity in various collision centralities in Au + Au collisions from $\sqrt{s_{NN}}$ = 7.7 - 200 GeV. The data points are compared to the Poisson expectation values and the bottom panel shows the ratio of $S\sigma$ over the Poisson expectation. Uncertainties included are statistical (bars) and systematical (brackets).}
\label{highmoments}
\end{minipage}
\end{figure}

Directed flow ($v_1$) of protons has been proposed as a sensitive probe to the softening of the equation of state and/or the first order phase transition. STAR made systematic measurements of $v_1$ of identified particles~\cite{Pandit}. The rapidity dependence of $v_1$, often quantified as the slope parameter $dv_1/dy'$ ($y'\equiv y/y_{\rm{beam}}$) vs. collision energy in Au + Au collisions from 10-40\% centrality are plotted in Fig.~\ref{proton_v1slope}. We observe that the proton $v_1$ slope changes sign from positive to negative when the collision energy increases from 7.7 GeV to 11.5 GeV, and it rises with increasing energy, but stays negative and approaches zero at 200 GeV. The $v_1$ slopes of other particles ($\bar{p}$, $\pi^{\pm}$, $K^{\pm}$) are all negative in our measured energy region. In the bottom panel of Fig.~\ref{proton_v1slope}, we take the difference in the slope parameters between protons and anti-protons weighted by their relative production yields, $i.e.$ ``net-proton" $v_1$ slope. A striking observation is that the net-proton $v_1$ changes sign twice in the measured energy region, and shows a minimum between 11.5-19.6 GeV. Also shown on the figure are transport model calculations. However, neither of these calculations can reproduce the observed net-proton $v_1$ slope. Other physics sources are under investigation.

System at the QCD critical point region is expected to show a sharp increase in the correlation length, thus induce large non-statistical fluctuation. Higher moments of conserved quantities (e.g. net-baryon number) are proposed to be an ideal probe as they are more sensitive to the critical point induced fluctuations~\cite{Stephanov}. A non-monotonic behavior of high moment distributions vs. collision energy is expected to be a signature of the QCD critical point. In Fig.~\ref{highmoments}, we present the STAR measurements of moment products $S\sigma$ and $\kappa\sigma^2$ of net-proton multiplicity distributions in various collision centralities in Au + Au collisions from 7.7 to 200 GeV~\cite{Luo}. Also plotted in the figure are the Poisson expectations in the first two panels and the ratio of $S\sigma$ over the Poisson expectation is shown in the bottom panel. We observe that in 0-5\% central collisions, moment products deviate from Poisson expectation at $\sqrt{s_{NN}}>$ 7.7 GeV, while in peripheral collisions, the data points are above the Poisson expectation below 19.6 GeV. The UrQMD transport model calculations show a monotonic behavior with collision energy. Data points below 19.6 GeV have large uncertainties which prevents us from drawing a conclusion on the energy dependence behavior. Details of other fluctuation measurements can be found in~\cite{ChenMcDonaldTribedy}.

STAR has completed the BES phase-I. We have observed many significantly different features compared to top RHIC energy. To locate the QCD phase boundary and the critical point, we need more precise measurements focusing at the energy region $\sqrt{s_{NN}}<\sim$ 20 GeV.

%

\section{Outlook}

STAR has just entered the era of precision QCD measurements. In the next decade, we have outlined a set of fruitful and compelling science programs with upgrades on both the detector subsystems and the RHIC machine~\cite{HHuang}. In the near term, STAR has two major subsystem upgrades - the Heavy Flavor Tracker and the Muon Telescope Detector - targeting for precision measurements on heavy flavors and dileptons to measure the sQGP properties. The current exciting but inconclusive BES results call for the BES phase-II program of precision measurements to map out the QCD phase structure. In the long term, STAR is exploring the upgrades at forward/backward regions to expand our physics program in $p$($e$)-A collisions to precisely study QCD in the cold nuclear matter.

\end{document}